\documentclass[aps,prb,twocolumn,superscriptaddress]{revtex4-1}

\usepackage{amsmath,amssymb}
\usepackage{graphicx}
\usepackage{epstopdf}
\usepackage{bm}
\usepackage{color}
\usepackage{mhchem}

\newcommand{\lbcoa}     {La$_{1.875}$Ba$_{0.125}$CuO$_4$}

\newcommand{\lesco}    {La$_{1.8-x}$Eu$_{0.2}$Sr$_x$CuO$_4$}

\newcommand{\lmsco}    {La$_{2-x-y}$M$_y$Sr$_x$CuO$_4$}

\newcommand{\tlt}      {$T_\text{LT}$}
\newcommand{\tht}      {$T_\text{HT}$}
\newcommand{\tco}      {$T_\text{CO}$}
\newcommand{\tso}      {$T_\text{SO}$}
\newcommand{\hsf}      {$H_\text{sf}$}

\newcommand{\la}       {$^{139}$La}

\newcommand{\slr}      {$T_1^{-1}$}
\newcommand{\ltot}     {LTO $\rightarrow$ LTT}
\newcommand{\htot}     {HTT $\rightarrow$ LTO}

\begin{document}

\title{Magnetic field induced anisotropy of $^{139}$La spin-lattice relaxation rates in 
stripe ordered La$_{1.875}$Ba$_{0.125}$CuO$_4$}  

\author{S.-H. Baek}
\email[]{sbaek.fu@gmail.com}
\affiliation{IFW-Dresden, Institute for Solid State Research,
PF 270116, 01171 Dresden, Germany}
\author{Y. Utz}
\affiliation{IFW-Dresden, Institute for Solid State Research, PF
270116, 01171 Dresden, Germany}
\affiliation{Institut f\"ur Festk\"orperphysik, Technische Universit\"at
Dresden, 01062 Dresden, Germany}
\author{M. H\"ucker}
\affiliation{Condensed Matter Physics and Materials Science Department, 
Brookhaven National Laboratory, Upton, New York 11973, USA} 
\author{G. D. Gu}
\affiliation{Condensed Matter Physics and Materials Science Department, 
Brookhaven National Laboratory, Upton, New York 11973, USA}
\author{B. B\"uchner}
\affiliation{IFW-Dresden, Institute for Solid State Research,
PF 270116, 01171 Dresden, Germany}
\affiliation{Institut f\"ur Festk\"orperphysik, Technische Universit\"at
Dresden, 01062 Dresden, Germany}
\author{H.-J. Grafe}
\affiliation{IFW-Dresden, Institute for Solid State Research, PF
270116, 01171 Dresden, Germany}

\date{\today}

\begin{abstract}

We report $^{139}$La nuclear magnetic resonance studies performed on a 
\lbcoa\ single crystal. The data show that the structural phase 
transitions (high-temperature tetragonal $\rightarrow$ low-temperature 
orthorhombic $\rightarrow$ low-temperature tetragonal phase) are 
of the displacive type in this material. The \la\ spin-lattice relaxation rate 
\slr\ sharply upturns at the charge-ordering temperature \tco\ = 54 K, 
indicating that charge order triggers the slowing down of spin fluctuations. 
Detailed temperature and field dependencies of the \slr\ below the 
spin-ordering temperature \tso\ = 40 K reveal the development of enhanced spin  
fluctuations in the spin-ordered state for $H \parallel [001]$, which are 
completely suppressed for large fields along the CuO$_2$ planes. Our results 
shed light on the unusual spin fluctuations in the charge and spin stripe 
ordered lanthanum cuprates.     


\end{abstract}


\maketitle

Numerous diffraction experiments have established the unidirectional spin/charge 
stripe model\cite{tranquada95,fujita02,hucker07,fink09,croft14,christensen14,thampy14}  
in the single-layer lanthanum-based cuprates, La$_{2-x}$(Ba,Sr)$_x$CuO$_4$ and 
\lmsco\ (M = Nd, Eu). The simple stripe picture, however, misses the leading 
electronic instability of stripe order and its relation to superconductivity. 
For example, it is largely unclear how charge order preceding spin order 
evolves to uniaxially modulated charge/spin stripe order. 


X-ray diffraction experiments in high magnetic fields have shown that charge 
order is enhanced when superconductivity is suppressed by the magnetic field. 
However, in 1/8 doped \lbcoa\, where the stripe order is most stable and 
bulk superconductivity is absent already in the zero field, high magnetic fields 
have little effect on the charge order.\cite{hucker13} Not much is known 
about anisotropic effects of magnetic fields applied along [001] and [100]. 
Measurements of the static susceptibility indicate that the spin order is 
stabilized for high magnetic fields $H \parallel [100]$. Furthermore, a spin 
flop occurs at a magnetic field $H \geq 6$ T along this direction.\cite{hucker08} 
Nuclear magnetic resonance (NMR) evidences unusual glassy spin 
fluctuations (SFs) below the spin-ordering 
temperature,\cite{hunt99,suh00,curro00,hunt01,simovic03,mitrovic08} but whether 
these spin  
fluctuations are related to charge order and whether there are anisotropy effects 
are not known.       

Another issue of current research is the coupling of the charge order to the 
lattice. It is widely believed that the low-temperature orthorhombic (LTO) 
$\rightarrow$ low-temperature tetragonal (LTT) structural phase transition  
has a profound effect on the stabilization of static charge/spin order which 
then suppresses superconductivity.\cite{buchner94,klauss00} Recent studies 
show that long range LTT ordering may not be essential for stripe order, but 
that local distortions may be enough to pin charge 
order.\cite{hucker10,guguchia13,christensen14,croft14,fausti11,forst14} This  
indicates that the coupling mechanism among the lattice, spin/charge stripes, 
and superconductivity is far more complex and remains to be fully understood.

NMR is an ideal technique to investigate such a complex coupling mechanism, 
because it probes the local  
spin/charge environment surrounding a nucleus, in particular, low-frequency 
spin fluctuations associated with various phase transitions. Since the $^{63}$Cu 
is too strongly influenced by the Cu moments leading to the wipeout of the NMR 
signal,\cite{hunt99,curro00,teitelbaum00} the \la\ nucleus is better suited 
to investigate the stripe phase and the structural phase transitions 
(SPTs).\cite{suh00,simovic03,grafe10,baek12a,baek13a}

Here, we show by means of \la\ NMR that additional spin fluctuations develop 
in the spin-ordered state. These SFs are strongly anisotropic in large 
magnetic fields: applied along the CuO$_2$ planes, the large magnetic fields 
lead to a suppression of these additional SFs, and static hyperfine fields 
lead to a broadening and loss of the \la\ signal intensity. In contrast, 
magnetic fields perpendicular to the CuO$_2$ planes have a weak effect on the 
spin fluctuations, and an additional relaxation mechanism enhances the \la\ 
nuclear spin-lattice relaxation. The observed anisotropy goes along with the 
enhanced spin order for a magnetic field parallel to the CuO$_2$ 
planes.\cite{hucker08} Our experiments also allowed for a detailed look at 
the local  
crystal structure, which has been the subject of a long debate. We find that 
structural phase transitions in \lbcoa\ are of the displacive type, locally 
probing the average structure given by diffraction studies.

The \lbcoa\ (LBCO:1/8) single crystal was grown with 
the traveling solvent floating zone  
method described in Ref.~\onlinecite{gu06}. The sample was accurately aligned along the 
magnetic field using a goniometer.  
\la\ (nuclear spin $I=7/2$) NMR spectra
were obtained by sweeping 
frequency at a fixed external field ($H$) in the temperature 
($T$) range 10 --- 300 K.\cite{clark95} Spin-echo signals each were taken by shifting 
50 kHz and their Fourier-transformed spectra were 
summed up to give rise to the full spectrum. Since the  
range of the sweeping frequency is quite narrow, i.e., much less than 2\% of the Larmor 
frequency, the frequency correction was not made. 
The spin-lattice relaxation rates \slr\ were measured
at the central transition ($+1/2 \leftrightarrow -1/2$) of \la\  
by monitoring the recovery of the echo signal after a saturating 
single $\pi/2$ pulse which, depending on the experimental conditions, ranges from 2 to 4 
$\mu$s. When the \la\ spectral width becomes broader at low temperatures, we 
carefully carried out the $T_1$ measurements to avoid any spectral diffusion. 
Then the following formula was used to fit the relaxation data to obtain \slr:
\begin{equation}
\begin{split}
\label{eq:T1}
1-\frac{M(t)}{M(\infty)}&=
a\left(\frac{1}{84}e^{-(t/T_1)^\beta}+\frac{3}{44}e^{-(6t/T_1)^\beta}\right.   \\
+ &\left.\frac{75}{364}e^{-(15t/T_1)^\beta}+\frac{1225}{1716}e^{-(28t/T_1)^\beta}\right),
\end{split}
\end{equation}
where $M$ is the nuclear magnetization and $a$ is a fitting parameter that is
ideally one.  $\beta$ is the stretching exponent, which becomes less than unity when 
\slr\ is spatially distributed, for example, in a spin glass.\cite{mitrovic08, narath67a} 

Figure 1 shows the \la\ NMR central transition 
($1/2\leftrightarrow -1/2$) as 
a function of $T$ obtained at $H=10.7$ T applied along the crystallographic 
directions [001] and [100] of the high-temperature tetragonal (HTT) unit cell. 
Different colors of spectra  
denote different structural phases.  For $H\parallel [001]$, the   
\la\ NMR line is quite narrow (full width at half maximum $\sim$30 kHz) and 
almost independent of $T$ in the HTT  
phase. Just below the \htot\ transition temperature \tht, the \la\ 
line undergoes an anomalous change and, upon further cooling, continues to 
broaden and shift to lower frequency. 
The $T$ evolution of the \la\ 
resonance frequency $\nu_0$ through \tht\ can be explained in terms of a  
second-order quadrupole shift 
which depends on the angle between the principal
axis of the electric field gradient (i.e., the axis of $V_{zz}$) and $H$:
\begin{equation}
\begin{split}
	\nu_0 &= \gamma_n (1+\mathcal{K})H \\
	      &+
	 \frac{15\nu_Q^2}{16\nu_0}\left[1
-\cos^2(\theta\pm\alpha)\right]\left[1-9\cos^2(\theta\pm\alpha)\right], 
\end{split}
\end{equation}
where $\gamma_n$ is the 
nuclear gyromagnetic ratio, $\mathcal{K}$ is the Knight shift, $\nu_Q$ is the 
quadrupole frequency,   
$\theta$ is the angle between [001] and $H$, and $\alpha$ is the tilt angle of the 
CuO$_6$ octahedra with respect to [001].  
Since $\theta=0$ for $H\parallel [001]$ and $\mathcal{K}$ for \la\ is 
very small in La-based cuprates,\cite{baek12a} the abrupt decrease of 
$\nu_0$ below \tht\  
indicates that $\alpha$ of the second term in Eq. (2) becomes non-zero 
and gradually increases with  
decreasing $T$ in the LTO phase.  The much larger linewidth in the LTT phase 
is then ascribed to local tilt disorder, i.e., a spatial distribution of $\alpha$. 

\begin{figure}
\centering
\includegraphics[width=\linewidth]{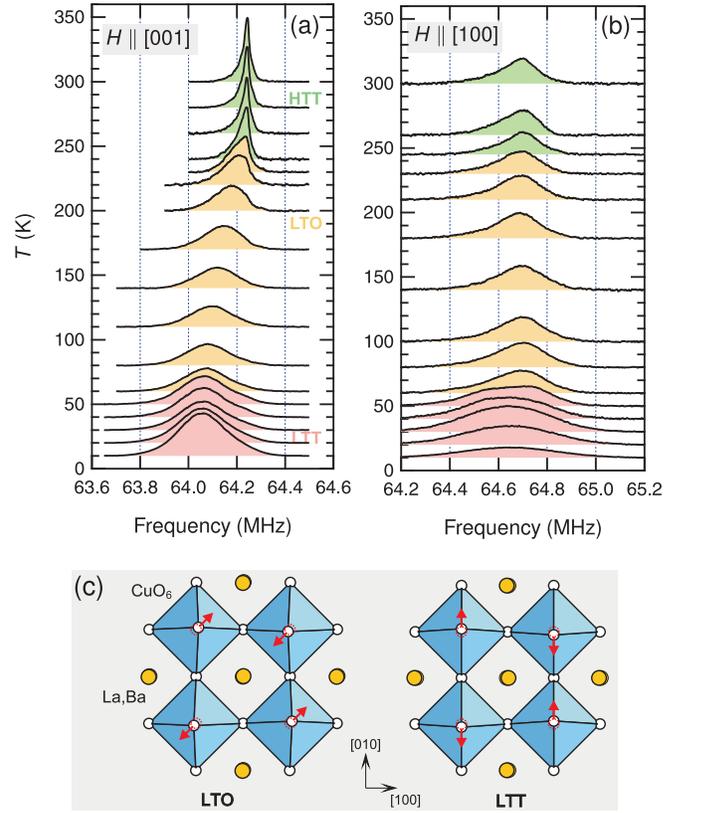}
\caption{\label{fig:la} $^{139}$La NMR central transition spectrum as a 
function of $T$ at 10.7 T applied along (a) [001] and (b) 
[100] in the HTT setting. The anomalous changes of the \la\ spectrum were 
observed at \tht\ for $H\parallel [001]$ and at \tlt\ for $H\parallel [100]$, 
respectively. Below \tso, the signal intensity becomes strongly anisotropic at low $T$.
Both Boltzmann- and $T_2$-corrections were made for the signal intensities.  
(c) Top view of LTO and LTT phases with one CuO$_2$ plane in the average structure of 
doped La$_2$CuO$_4$. The arrows denote the tilt direction of the CuO$_6$ 
octahedra. 
}
\end{figure}

This is strong evidence that the \htot\ transition can be described as a 
transition from flat CuO$_2$ planes in the HTT phase to a phase with tilted 
CuO$_6$ octahedra.\cite{baek12a} Note that we do not observe LTT-type tilt 
fluctuations persisting through the transitions, as has been found recently by 
a combination of neutron powder diffraction and inelastic neutron scattering,\cite{bozin15} 
most likely due to the different time scales of neutron-scattering and NMR 
experiments. For NMR linewidth measurements, fluctuations on a 10$^{-2}$ ms 
timescale are already enough to average out the effects on the resonance 
lines.  

However, our results are in agreement with the average structure from 
conventional diffraction studies,\cite{katano93,axe94,friedrich96,braden01} 
rather than the \textit{local structure} model which proposes an order-disorder type 
transition.\cite{billinge94,haskel96,haskel00,han02,wakimoto06} In the average structure model there is no tilt of the CuO$_6$ octahedra in the HTT phase, and the \htot\ transition 
is determined basically by the tilt angle of the octahedra. At the \ltot\ 
transition, the tilt axis rotates by $45^\circ$ in this model. On the other 
hand, in the local structure model the LTO structure is built up from the coherent 
spatial superposition of local LTT structures. Here, the tilt axis does not 
rotate at the \ltot\ transition, and does not vanish at the \htot\ transition, 
which is not consistent with our data. Note, however,  that for 
$H\parallel [100]$ ($\theta=90^\circ$), the quadrupole broadening is very 
large and obscures the tilting effect at \tht. On the other hand, through the 
\ltot\ transition at \tlt, only for $H\parallel [100]$ is a clear anomaly 
observed. This observation as well is consistent with the average structure 
model. As illustrated in Fig. 1(c), when $H \parallel [100]$, the rotation of 
the octahedral tilt direction below \tlt\ should lead to a change of the  
direction of  $V_{zz}$ with respect to $H$. This is different from the case of 
$H\parallel [001]$ where $\alpha$ remains the same. We conclude that all 
structural phase transitions in Ba-doped La$_2$CuO$_4$ are in agreement with 
the average structure model.

The dynamic properties of the structural phase transitions and, in particular, 
of the spin fluctuations in the stripe ordered phase can be probed by the \la\ 
spin lattice relaxation rate \slr. Figure 2(a) shows \slr\  
as a function of $T$ at $H=10.7$ T applied along [001] and [100], revealing 
sharp anomalies at \tht\ and \tlt\ regardless of the field orientation. While 
the sharp peak at \tht\ represents the thermodynamic critical mode associated 
with the \htot\ transition, the rapid upturn at \tlt\ is most likely not 
caused by the \ltot\ transition itself, because \slr\ is expected to 
\textit{drop} sharply below \tlt, as was detected in \lesco\ 
(LESCO).\cite{curro00,baek13a} Indeed, a close look at Fig. 2(b) implies that 
the  
rapid upturn of \slr\ by up to three orders of magnitude is most likely due to 
the spin ordering at 40 K. The \slr\ upturn starts at the charge ordering 
temperature\cite{hucker11} $T_\text{CO}\sim T_\text{LT}\sim 54$ K  suggesting 
that the charge ordering triggers the critical slowing down of SFs toward spin 
ordering.\cite{hunt01,julien01}
\begin{figure}
\centering
\includegraphics[width=0.9\linewidth]{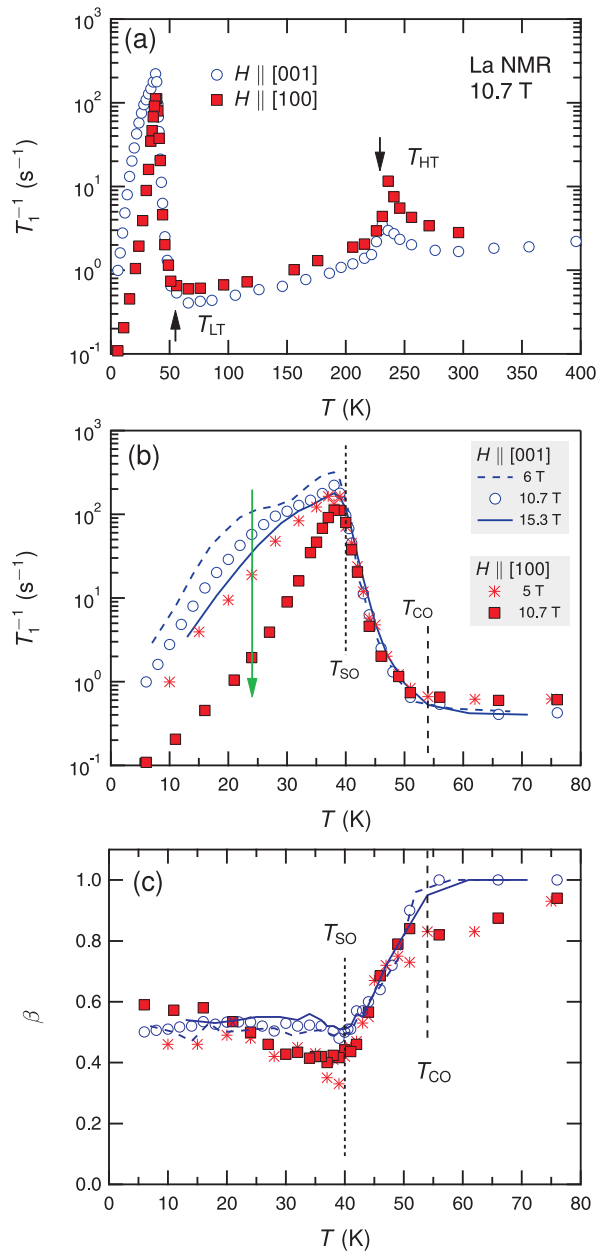}
\caption{\label{fig:T1} (a) $T$ dependence of \la\ \slr\ at 
10.7 T applied along [001] and [100].  (b) \slr\ vs $T$ at various magnetic 
fields $H$.   
The onset of the \slr\ upturn coincides with \tco\ independent of $H$. Only 
below \tso\ is the strong dependence of \slr\ on the strength and orientation of $H$ 
observed. The green arrow indicates the temperature where a detailed field 
dependence has been measured. 
(c) Stretching exponent $\beta$ as a function of $T$ and  $H$, which correlated with \slr.    
}
\end{figure}

Further, in Fig. 2(b) the field dependence of \slr\ reveals interesting 
features in the stripe phase. In the temperature range 
$T_\text{SO}<T\leq T_\text{CO}$, despite the huge enhancement of \slr\ by more 
than three orders of magnitude, $T_1^{-1}(T)$ is independent of orientation 
and strength of $H$. This suggests that the spin fluctuations are still 
isotropic and independent of $H$ above \tso , consistent with spin 
fluctuations of a two-dimensional (2D) quantum Heisenberg antiferromagnet or an effective 
spin-liquid state.\cite{tranquada08} On the other hand, the static 
susceptibility indicates that the spin dimensionality is already effectively reduced 
from 2D Heisenberg to 
2D XY below \tco.\cite{hucker08} However, once the spins are 
ordered below \tso\ = 40 K, $T_1^{-1}(T)$ also reveals a strongly anisotropic 
field dependence.  

At 10.7 T $\parallel [100]$, $T_1^{-1}(T)$ displays the expected behavior for 
slow spin dynamics driven by conventional antiferromagnetic (AFM) correlations with a glassy 
nature: on the low temperature side, \slr\ decreases steeply  
consistent with a conventional Bloombergen, Purcell, and Pound (BPP) 
mechanism.\cite{suh00,curro00} In contrast, for small fields parallel to [100] 
as well as  
for all studied fields $H \parallel [001]$, the relaxation rate remains 
significantly enhanced below \tso . This enhanced relaxation has already been 
observed in the stripe ordered phase of L(E)SCO, and led the authors to modify 
or even abandon the simple BPP model.\cite{suh00,simovic03,curro00,mitrovic08} 
Our results show that the spin-lattice relaxation deviates from the simple BPP 
model for low fields and for  
$H \parallel [001]$. A possible reason for this deviation is that the field 
along the planes stabilizes the spin order.\cite{hucker08} Small fields or a 
field perpendicular to the planes allow for the peculiar spin fluctuations 
that lead to the enhanced spin lattice relaxation and deviation from the 
simple BPP model below \tso, as will be discussed in detail below. 

Another fingerprint of glassy spin dynamics besides the (modified) BPP 
behavior, and a measure of a distribution of spin-lattice relaxation rates, is 
a stretching exponent $\beta$ that deviates from one [see Eq. (1)]. $\beta$ is 
presented in Fig. 2(c) as a function of $H$ and $T$, and exhibits distinct 
changes at \tco\ and \tso, which correlate with $T_1^{-1}(T)$. The decrease of 
$\beta$ below \tco\ indicates that the charge ordering initiates the 
distribution of $T_1^{-1}$, and therefore of the inhomogeneous spin 
fluctuations. Below \tso, $\beta(T)$ is weakly $T$ and $H$ dependent, i.e. a 
large, but $T$-independent distribution of spin fluctuations is still 
present.\cite{mitrovic08} The anisotropic behavior of $T_1^{-1}$ below \tso\  
is, however, not reflected in the distribution of spin lattice relaxation rates.

Since the multi-exponential relaxation function [Eq. (1)] is complicated and 
the values of \slr\ obtained from a stretched fit are not the average \slr, 
\cite{mitrovic08} we show in Fig. 3 typical recovery curves and fits for $T$ 
= 16, 42, and 66 K. Clearly, a stretching exponent $\beta$ is needed to 
account for the distribution of spin-lattice relaxation rates. On the other 
hand, the values of \slr\ fitted with or without the stretching exponent do 
not deviate substantially: At 66 K, $T_1$ = 1677 ms with $\beta$ = 0.87, and 
$T_1$ = 1660 ms ($\beta$ = 1). At 16 K $T_1$ = 2208 ms with $\beta$ = 0.58, and 
$T_1$ = 2194 ms ($\beta$ = 1). For fast relaxation at 42 K, the deviations of 
\slr\ depending on the stretching exponent are larger: $T_1$ = 49 ms ($\beta$ 
= 0.46) and $T_1$ = 72 ms ($\beta$ = 1). However, when plotting \slr\ on a 
log scale as in Fig. 2(b), the deviation is hardly larger than the point size 
for 42 K. Therefore, the stretched relaxation has no impact on the main 
findings of our work. 

\begin{figure}
\centering
\includegraphics[width=0.9\linewidth]{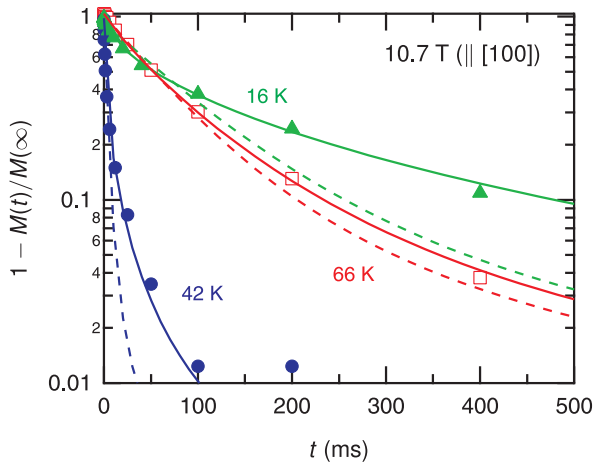}
\caption{\label{fig:Relax} Relaxation curves of nuclear magnetization $M$ for 
$T$ = 16, 42, and 66 K at 10.7 T parallel to [100]. The 
solid and dashed lines are fits by Eq. (1) with $\beta$ 
as a free parameter and with $\beta=1$, respectively. See text for details.} 
\end{figure}         

In order to gain a better understanding of the anisotropic SFs, we examined in 
detail the field dependence of \slr\ at a fixed temperature of 24 K, which is 
shown in Fig. 4. Figure 2(b) already revealed that $T_1^{-1}(T)$ is strongly 
suppressed with increasing $H\parallel [100]$ from 5 to 10.7 T, while changes 
for $H\parallel [001]$ are much weaker. Figure 4 further verifies that \slr\ for 
$H\parallel [100]$ is reduced much faster than that for $H\parallel [001]$ 
with increasing $H$, and thus the \slr\ anisotropy increases accordingly. As 
expected, the dashed and dotted lines in Fig. 4 indicate that the \slr\ is 
almost isotropic for H=0. For a quantitative understanding of the anisotropic 
spin fluctuations, it is convenient to define new spin-lattice relaxation 
rates:   
$R_i \equiv T\gamma_n^2\sum_\mathbf{q} A_i^2 \chi''_i (\mathbf{q},\omega_0)/\omega_n$
, where $i=a,b,c$ represents one of the crystallographic axes, $\chi''$ is the 
imaginary part of the dynamical susceptibility, and $A_i$ is the hyperfine 
coupling constant.\cite{baek10c} This notation emphasizes the fact that \slr\ probes only 
the SFs perpendicular to the nuclear quantization axis, i.e. 
$(T_1^{-1})_{[001]}=R_a+R_b$ and $(T_1^{-1})_{[100]}=R_b+R_c$ for a given 
temperature. Above \tso, our data indicate that $R_a=R_b=R_c$, i.e., isotropic 
hyperfine coupling and  
Heisenberg-type SFs. Now, let us take two \slr\ values at 24 K and 10.7 T 
where \slr\ is   
different by more than an order of magnitude for the two different field 
orientations [see Fig. 2(b)]. Then, we have  
$(R_a+R_b)_{[001]} \approx 10 (R_b+R_c)_{[100]}$. With $R_a=R_b$ due to the 
macroscopic tetragonal symmetry with $H \parallel [001]$, we get 
$2 (R_b)_{[001]} \approx 10 (R_b+R_c)_{[100]}$. Therefore, no matter how small  
$R_c$ may be, $(R_b)_{[001]} \gg (R_b)_{[100]}$. In other words, a field 
parallel to [100] strongly suppresses all SFs, whereas the spin fluctuations 
parallel to the CuO$_2$ planes are not affected for $H\parallel [001]$. 
This is because the spins are confined to the CuO$_2$ planes at least below 
\tso. Due to the strong AFM coupling, the spins orient perpendicular to the 
external magnetic field. For $H\parallel [001]$, they are already 
perpendicular, and thus the fluctuations parallel to the planes are not affected 
by $H\parallel [001]$, and \slr\ is enhanced. In contrast, a field 
$H\parallel [100]$ creates an in-plane anisotropy that tends to align the 
spins perpendicular to the field. Now, the spins cannot fluctuate as freely 
within the planes as for the zero field or as for $H \parallel [001]$, and the 
larger the applied magnetic field, the stronger is this effect. 

Interestingly, we observed a small but clear anomaly at \hsf\ $\approx$ 7 T 
for $H\parallel [100]$, which is attributed to the spin-flop 
transition.\cite{hucker08} In the simple stripe picture, the direction of 
spins   
alternates between [100] and [010] in neighboring planes owing to the coupling 
to the LTT structure. For $H\parallel [100]$, spins along [010] are further 
stabilized, but those along [100] at first are destabilized when the field 
becomes of the order of the in-plane spin-wave gap. The consequence is a 
spin-flop transition at $H$ = \hsf\ where these spins change their direction 
from  [100] to [010].\cite{hucker08} Right at \hsf\ $\approx$ 7 T, we indeed 
observe a local maximum in \slr\ for $H \parallel [100]$, which reflects the 
enhanced fluctuations of the destabilized spin sublattice. Upon further 
increasing  $H > H_{sf}$, \slr\ decreases rapidly again reflecting the 
stabilized spin order, and indicating that now these spins are also stabilized 
in an in-plane direction perpendicular to the field.    

\begin{figure}
\centering
\includegraphics[width=0.9\linewidth]{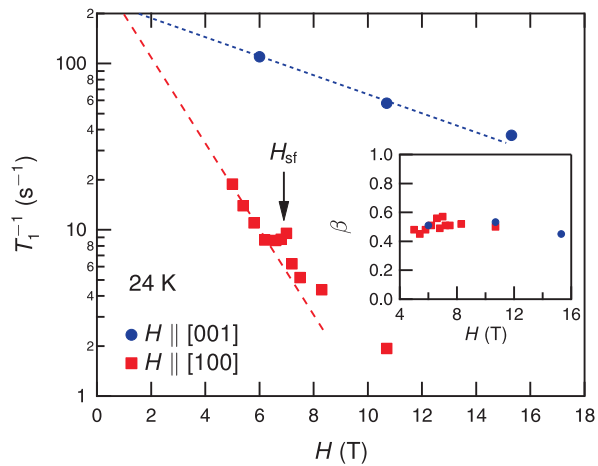}
\caption{\label{fig:T1} Detailed field   
dependence taken at a fixed  
temperature of 24 K [see the green vertical arrow in Fig. 2(b)] reveals that the 
\slr\ anisotropy rapidly 
increases with increasing $H$. Small anomaly at $\sim 7$ T is  
ascribed to the spin flop transition. The dotted and dashed lines are guides 
to the eye. The inset reveals that $\beta$ is almost independent of $H$.   
}
\end{figure}

Further evidence for a stabilization of the spin order for large fields 
parallel to [100] is provided by the strong anisotropy of the \la\ signal 
intensity below the spin-ordering temperature \tso. As can be seen in Fig. 1, 
the integrated NMR signal intensity $I_\text{int}$ for $H\parallel [100]$ is rapidly 
reduced below \tso, in stark contrast to that for $H\parallel [001]$ 
which is constant or even appears to increase at low temperatures. 
Whereas the NMR intensity can be easily affected by the 
temperature-dependent gain arising from, e.g., the change of the $Q$ factor of the 
NMR circuit, the relative intensity at a given temperature should not.  
Therefore, the clearly different temperature dependence of $I_\text{int}$ for 
the two field orientations evidences the strong anisotropy of 
$I_\text{int}$ at low temperatures. 
Since the enhancement of the \la\ signal intensity is unlikely intrinsic, the  
strong anisotropy is ascribed to the loss of $I_\text{int}$ for $H\parallel [100]$.   
This rapidly disappearing \la\ signal intensity looks similar to the wipeout of 
the $^{63}$Cu spectra,\cite{hunt99,hunt01,curro00} which is caused by a dramatic shortening 
of the relaxation times  
($T_2$ and $T_1$) due to a high spectral density of electronic fluctuations at 
the Larmor frequency.\cite{julien01} While this wipeout effect is not known to 
depend on the field orientation, the loss of the \la\ signal intensity 
for $H\parallel [100]$ differs from that of the $^{63}$Cu spectra and may be 
caused by static internal  
hyperfine fields that mainly shift, and, due to a distribution of hyperfine 
fields, may also spread the \la\ intensity over a broad frequency range. 
On the other hand, the significantly larger \slr\ below \tso\ for 
$H \parallel [001]$ [see Fig. 2(b)] indicates the persistence of strong 
spin fluctuations, which could induce incomplete spin ordering in this field 
direction. Then this naturally accounts for the strongly anisotropic \la\ signal  
intensity below \tso. 

In summary, our NMR results reveal the displacive type of all structural phase 
transitions in \lbcoa\ and that the local structure is compatible with the 
average structure determined by diffraction experiments. The slowing down of 
AFM spin fluctuations below the \ltot\ transition is triggered by 
the concomitant onset of charge order. Below the spin ordering temperature, 
\tso, we observed a strong anisotropy of the spin-lattice relaxation rate at large 
fields. With increasing field, the spin fluctuations are rapidly suppressed 
for $H\parallel [100]$, while they are weakly suppressed for $H\parallel [001]$.  
We conclude that the spin order is stabilized at large fields only for 
$H\parallel [100]$ involving the spin flop transition at 
$\sim 7$ T $\parallel [100]$.   Our results 
resolve the reason for the deviations from the simple BPP model below the 
spin-ordering temperature.

This work has been supported by the DFG Research Grant No. BA 4927/1-1.
M.H. acknowledges support by the Office of Science, U.S. Department of Energy
under Contract No.~DE-AC02-98CH10886.

\bibliography{mybib}

\end{document}